\documentclass[reprint,aps,prstper]{revtex4-1}

\usepackage{graphicx}% Include figure files
\usepackage[export]{adjustbox}% to put frame around figure
\usepackage{hyperref}% add hypertext capabilities
\usepackage{footnote}
\usepackage{amsmath,amssymb}
\usepackage{enumitem}
\usepackage{lineno}
\begin{document}

\title{Epistemic stances toward group work in learning physics: Interactions between epistemology and social dynamics in a collaborative problem solving context}

\author{Jessica R. Hoehn}
\email[]{jessica.hoehn@colorado.edu}
\affiliation{Department of Physics, University of Colorado, 390 UCB, Boulder, Colorado 80309, USA}

\author{Julian D. Gifford}
\affiliation{Department of Physics, University of Colorado, 390 UCB, Boulder, Colorado 80309, USA}

\author{Noah D. Finkelstein}
\affiliation{Department of Physics, University of Colorado, 390 UCB, Boulder, Colorado 80309, USA}

\begin{abstract}
As educators we often ask our physics students to work in groups---on tutorials, during in-class discussions, and on homework, projects, or exams. Researchers have documented the benefits of group work for students’ conceptual mastery and problem solving skills, and have worked to optimize the productivity of group work by assigning roles and composing groups based on performance levels or gender. However, it is less common for us as a physics education research community to attend to the social dynamics and interactions among students within a collaborative setting, or to address students’ views about group work. In this paper, we define \textit{epistemic stances toward group work}: stances towards what it means to generate and apply knowledge in a group. Through a case study analysis of a collaborative problem solving session among four physics students, we investigate how epistemic stances toward group work interact with social dynamics. We find that misalignment of stances between students can inform, and be informed by, the social positioning of group members. Understanding these fine-grained interactions is one way to begin to understand how to support students in engaging in productive and equitable group work. 
\end{abstract}

\maketitle

\section{\label{sec:intro}Introduction}
With the increasing use of interactive engagement techniques in physics education, we are more frequently asking our students to work in groups (on tutorials~\cite{McDermottShaffer2002}, during class discussions~\cite{Mazur1997}, on laboratory or classroom-based projects~\cite{Dounas-FrazerStanleyLewandowski2017, Feder2017}, and sometimes even on exams~\cite{WiemanRiegerHeiner2014, CarrSaultWolf2018}). Interactive engagement, which often includes group work and collaboration, is cited time and again for being beneficial for student learning as measured by performance on conceptual assessments, problem solving assessments, or course grades~\cite{Hake1998, Chi2009, LumpeStaver1995, HellerKeithAnderson1992, FreemanEtal2014}. There are also many elements of student learning that we care about beyond conceptual understanding and content mastery. Interactive engagement, including group work, can help students develop positive attitudes towards science, and a sense of community, identity, and belonging~\cite{SpringerStanneDonovan1999, EtkinaEtal1999, Esmonde2009, ArcherEtal2017, Tonso2006, LewisEtal2017}. As physics educators, we care about group work not only because it benefits student learning, but because collaboration is a key element of science, and because collaborative environments can help facilitate positive experiences for students. However, the \textit{existence} of group work in our classrooms is not sufficient. We must also pay attention to how it is implemented and consider in what ways and for whom the group work is beneficial. 

The productivity of group work, both in terms of students’ learning and the creation of supportive and inclusive environments, hinges on the social dynamics among the participants. In a group work setting, students’ social identities become increasingly relevant~\cite{CooperBrownell2016}; despite the many benefits that group work provides, interactions in these collaborative settings may also lead to exclusive or inequitable environments~\cite{CooperBrownell2016, CooperDowningBrownell2018}. Some researchers have found that groups are more effective when they are intentionally formed to include mixed “ability” (as measured by individual exam scores)~\cite{HellerHollabaugh1992}, or to avoid mixed gender groups that are male-dominated~\cite{GroverItoPark2017, HellerHollabaugh1992}, while others suggest that it is best to have students self assemble into groups so that they can work with people they are most comfortable with~\cite{CooperDowningBrownell2018}. Regardless of how groups are formed, students’ views about the collaborative nature of science~\cite{QuanElby2016} or their framing of the type of activity they are engaged in~\cite{ScherrHammer2009, IrvingMartinukSayre2013} will also impact the social interactions among students and thus the productivity or effectiveness of the group work. Here, we attend to an epistemological aspect of group work as a mechanism for fostering (or undermining) equitable group participation.

In this paper, we present an analysis of a collaborative problem solving session in which we identify a new construct called \textit{epistemic stances toward group work}---views about how knowledge will be generated and applied in a group---and argue that the (mis)alignment of these stances among group members is connected to the social positioning of individuals within the group. We apply this idea to examine how one student guides most of the group’s sense making, yet is simultaneously positioned as less knowledgeable. We identify two different stances among four individual group members and find that as one stance dominates the other in the group sense making process, one student, whose behaviors are aligned with the non-dominant stance, is positioned as less knowledgeable. In this case, we argue that the misalignment of epistemic stances toward group work contributes to, and is reinforced by, this social positioning. 

The construct of epistemic stances toward group work brings together the ideas of epistemology~\cite{Hofer2001}, framing~\cite{Tannen1993}, group work~\cite{JohnsonJohnsonSmith1998}, and social dynamics~\cite{Barron2003,TheobaldEtal2017}, and is one way that we can begin to understand the participation and interaction among students in collaborative settings. The purpose of this paper is to introduce this construct and demonstrate its utility through a case study analysis. Identifying epistemic stances toward group work allows us to investigate the interactions between epistemology and social dynamics and better understand the social positioning of individual group members. In the next two sections we introduce the construct of epistemic stances toward group work and then discuss how it fits in with relative theoretical constructs in the physics education research community: implementation of group work, analysis or assessment of group work, epistemological framing, and productivity of group work.

\section{\label{sec:estgw}Epistemic stances toward group work}
For many physicists and physics students, what it means to know, learn, and do physics (their epistemology of physics) includes group work and collaboration. Likewise, what it looks like to engage in group work in a given context can be determined, in part, by individuals' beliefs about what counts as knowledge. Ideas about the form that group work should take and how it should function to generate knowledge may be different for different people. For example, in a given context, one person may think that group work should involve delegating individual tasks to individual people such that each group member contributes unique expertise or knowledge to the larger problem at hand, while another person may see group work as an opportunity for multiple people to work together simultaneously on one collective task. Or, one may view the goal of a group work scenario to be to reach an answer to a particular problem, while someone else may view group work as a time to collectively think through relevant ideas or principles such that individuals may later come to their own conclusions. These ideas about the form that group work can (or should) take are subsumed within the broader ideas of epistemology and group work. 

\begin{figure}
\includegraphics[scale=.5]{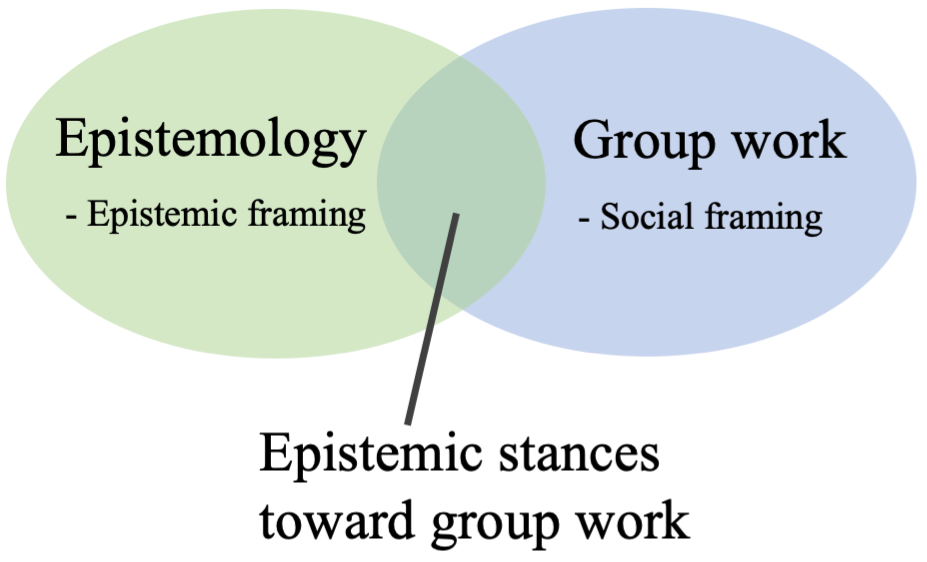}
\caption{The construct of epistemic stances toward group work sits at the intersection of epistemology and group work.}\label{fig:venndiagram}
\end{figure}

As shown in Fig. \ref{fig:venndiagram}, we consider the domains of epistemology and group work to be overlapping. Epistemology refers to beliefs about the nature of knowledge, while group work includes collaboration among individuals and is facilitated by social dynamics. We define \textit{epistemic stances toward group work} as views about how group work functions to generate and apply knowledge, existing at the intersection of these two domains (see Fig. \ref{fig:venndiagram}). Views about the role of group work in learning physics are subsumed by, but distinct from, views about learning physics more generally. Similarly, views or expectations about how group work should function exist as one element of the broader activity of group work, distinct from other contributing factors such as social norms or accountability. Epistemology and group work have each been the subject of extensive investigation within the physics education research community~\cite{Elby2010,AAAS_LeversForChange2019,NRC_DBER2012}, but we now look at the intersection. Zooming in on epistemological views about group work allows us to gain a more nuanced and rich understanding of students’ views about what it means to learn and do physics. Specifically, attending to students’ epistemic stances toward group work allows us to investigate the social dynamics and interactions among students that can support or inhibit productive and equitable group work. 

Much like epistemologies in general, epistemic stances toward group work may be fluid and context dependent~\cite{ElbyHammer2001, ChinnBucklandSamarapunga2011, Hofer2006, DreyfusHoehnEtal2019, MasonSingh2010, McCaskeyElby2005, LisingElby2005}. The stances people enact may vary moment to moment, and they are likely informed by a multitude of factors---interpersonal interactions, content domain, expectations set by the instructor, content and format of collaborative tasks, relationships among people in the group, etc. In identifying students’ epistemic ``stances’’, we refer to views that are enacted in the specific local context. These views may or may not be preferences or beliefs held by the students, and they may or may not be robust or permanent. We identify stances or views aligned with the students’ in the moment actions and behaviors in a collaborative problem solving context, but do not make claims about the students' preferences or stability of stances beyond the local context.

There are two different stances that we identify in our analysis. The first is the ``Collective Consensus Building’’ stance, in which group work entails generating and making sense of ideas collectively. Aligned with this view of group work, individuals contribute tentative ideas for the group to collectively negotiate and the sense making process is characterized by individuals thinking out loud and building off of each other's ideas. This type of collaboration has been described in prior research as ``co-construction''~\cite{LeupenKephartHodges2020, Chi2009, PawlakIrvingCaballero2018}. The second stance is ``Explainer-Explainee.’’ Here, group work means that individuals will come to understand the ideas at-hand and then explain them to others, a mutual process where if you understand something or know the answer, you then explain it to the group~\footnote{In the Explainer-Explainee stance, individuals reach an understanding of the content or topic at hand and then explain to others. This understanding may be reached individually (i.e., an individual learner works out a calculation on their own, recalls something they have previously learned, or consults a textbook or resource and makes sense of the content internally before sharing with others) or collectively (i.e., through group conversations, an individual comes to make sense of an idea, thanks to contributions or questions from peers). We do not tease apart this aspect of meaning-making in the Explainer-Explainee stance in our analysis, though we note that in both individual and collective meaning making, learning is a social activity.}. Although we identify only these two stances in our data, we do not claim that these are the only possible epistemic stances toward group work. We could imagine additional stances, and also note that these stances do not have to be fixed---there could be many stances or blends of stances that individuals or a group as a whole might take up. The goals of this paper are to introduce the construct of epistemic stances toward group work, identify the two above-mentioned stances as examples of the construct in the data, and explore how these stances (and the misalignment between them) can interact with the social dynamics and sense making of a group in a collaborative problem solving setting. 

\section{\label{sec:background}Background}
\subsection{\label{sec:implementation}Implementation of group work matters}
Active learning strategies, which often include group work, have been shown to be disproportionately beneficial for students from underrepresented groups, in terms of course grades~\cite{Preszler2009, HaakEtal2011}. Yet there is also evidence to suggest that active learning classroom interventions can have different impacts on different subpopulations of traditionally underrepresented students (e.g.,~\cite{EtkinaEtal1999, BeichnerEtal2007, EddyHogan2014}). Thus, while interactive engagement is generally thought to be “good” for student learning, we must investigate in what ways, and for whom, it is beneficial. 

Research in biology education identifies some of the ways in which group work, despite its potential benefits, can actually be isolating or harmful for some students. Cooper and Brownell~\cite{CooperBrownell2016} found that active learning strategies incorporating group work can increase the relevance of LGBTQIA social identities in the classroom, which can lead to stress about whether it is safe to come out, more opportunities for being misgendered, and increased cognitive load for students who are members of the LGBTQIA community. They state that “active learning changes the dynamics of the classroom so that who the instructors and students are has a larger impact on the student experience, particularly for students who are in the minority”~\cite[p. 3]{CooperBrownell2016}. In another study, Cooper, Downing, and Brownell~\cite{CooperDowningBrownell2018} explored the relationship between active learning practices and student anxiety, and found that group work had the potential to either increase or reduce student anxiety. The impact on students depended on how the group work was implemented (e.g., allowing students to choose their groups or giving students time to synthesize their thoughts before sharing with their peers tended to reduce students’ anxiety, whereas feeling uncomfortable with group members increased anxiety). Along the same vein, some researchers call on instructors and researchers to not only focus on implementing interactive engagement and group work, but to pay close attention to the dynamics among students in collaborative settings, highlighting the implications for equity and inclusion in our educational environments~\cite{EddyEtal2015, TraxlerEtal2016, Tonso2006}. In this paper, we respond to that call by investigating the social dynamics of one group of students in a collaborative problem solving setting and explore how the social dynamics are tied to epistemological stances toward group work to effect the differing social positioning of group members.

Tanner calls for instructors to design learning environments that attend to both individual students and interactions among students through teaching strategies that cultivate equitable classroom environments (e.g., establish classroom community norms, use varied active learning strategies, assign reporters for small groups)~\cite{Tanner2013}. Assigning and rotating through specific roles is one way to encourage equitable group discourse, and has been identified in physics as a way to promote effective group work~\cite{HellerHollabaugh1992}. Eddy \textit{et al.} call on instructors to structure their classroom activities in a way that promotes equity by understanding the differential participation and barriers to participation in peer discussions for some students~\cite{EddyEtal2015}. In a study of introductory biology students working in self-selected groups, they found that students' preferred roles in group work were correlated with their social identities (e.g., race, nationality, gender). The authors identified three barriers to participation in peer discussions: exclusion from discussion by group members, anxiety around participating in discussion, and low student perceptions of the value of group work~\cite{EddyEtal2015}. They argue that in order to promote equitable classroom environments, instructors must first understand the barriers that lead to differential participation in group work. In our case study analysis we identify misalignment of epistemic stances toward group work as a potential barrier to equitable participation in group problem solving.   

In line with these studies, other researchers have suggested the need to investigate factors that contribute to the construction and productivity of group work environments (e.g.,~\cite{PollockFinkelsteinKost2007, TurpenFinkelstein2010, Barron2003}). For example, while Lorenzo, Crouch, and Mazur report that use of interactive engagement reduces the gender gap on the Force Concept Inventory (FCI)~\cite{LorenzoCrouchMazur2006, HestenesWellsSwackhamer1992}, Pollock, Finkelstein, and Kost report that interactive engagement alone is not sufficient and that we must investigate the roles both instructors and students play in constructing norms around collaboration~\cite{PollockFinkelsteinKost2007}. Thus, in order to investigate the factors that allow social dynamics to support (or hinder) equitable discourse, we must attend to multiple dimensions of group work and the interactions among them.

\subsection{\label{sec:multiple-dimensions}Attending to multiple dimensions of group work}
Just as our community has documented that physics education is about more than content mastery, so too is group work more than assembling several individuals to a common task, or assigning roles to individual group members. In K-12 mathematics education, Barron illustrates through case study analyses how both cognitive and social factors play a role in defining a collaborative mathematics problem solving environment among sixth-grade students~\cite{Barron2000, Barron2003}. She identified more and less successful groups as defined by problem solving performance and uptake of correct ideas. ``Less successful'' groups had ``relational issues'', which included ``competitive interactions, differential efforts to collaborate, and self-focused problem solving trajectories''~\cite[pg. 348]{Barron2003} that impeded the group's ability to engage in productive problem solving. Barron suggests that one way the cognitive and social factors work together to construct a joint problem solving environment is through the ``development and maintenance of a between-person state of engagement''~\cite[pg. 349]{Barron2003}, which she describes as the awareness group members have for one another, ranging from ``complete lack of joint attention'' to ``continual coordinated participation.'' Along a similar vein, we attend to the ways in which social factors are intertwined with epistemology in the construction and dynamic evolution of a joint problem solving space. However, in considering the “success” or “productivity” of a group, we go beyond content mastery and correct answers. In our case study analysis, we consider the ways in which group problem solving interactions may be cohesive (or not) along social and epistemological dimensions, independent of conceptual productivity.  

Pawlak, Irving, and Caballero identified four modes of collaboration by attending to three dimensions of introductory physics students' interactions---social, discursive, and disciplinary~\cite{PawlakIrvingCaballero2018}. Along the social dimension, they characterized the overall tenor of students' collaboration as consonant or dissonant~\cite{LumpeStaver1995}. The discursive dimension identifies the interaction patterns among students: consensual (one student makes substantial contributions), responsive (multiple students make substantive contributions), elaborative (the substantive contributions from multiple students build off of one another)~\cite{HoganNastasiPressley1999}, and argumentation (involving evidence, a subsequent claim, and justification of how the evidence supports the claim)~\cite{Toulmin2003}. Regarding disciplinary content, they characterized students' conversations as specific or abstract. Treating the social, discursive, and disciplinary dimensions independently, Pawlak, Irving, and Caballero identified four distinct modes of collaboration---debate, informing, co-construction of an answer, and building understanding towards an answer---each arising from a unique combination of codes along the three dimensions~\cite{PawlakIrvingCaballero2018}. Similar to Pawlak \textit{et al.}'s study, we attend to multiple dimensions of students' collaboration, but our analysis differs because we consider the interactions between those dimensions and attend to finer-grained details of the social dynamics in a collaborative setting. 

Sohr, Gupta, and Elby also attend to students’ group work interactions by using multiple analytical lenses~\cite{SohrGuptaElby2018}. By investigating the intertwining of conceptual, epistemological, and socioemotional dynamics, they illustrate the multifaceted ways in which conflict can arise in collaborative settings, and identify one way students may resolve those conflicts. In particular, they describe an ``escape hatch'' as a series of discourse moves that serve to relieve tension in the group by ending or shifting the conversation before a conceptual resolution has been reached. On the surface, escape hatches (closing statements) can appear to be purely epistemological statements---not recognizing the complex intertwining of various factors might lead one to misinterpret these moves as an indicator of a group’s co-constructed epistemological stance rather than a result of, and resolution for, building tension within the group. Escape hatches can be productive in that they can help to establish more equitable group norms, or open up space for the group to engage collaboratively in conceptual discussion. This work provides an example of how interactional dynamics in a group can be the result of entanglement of many different factors. We build on this work by investigating the interactions between sense making~\cite{OddenRuss2019}, epistemology, and social dynamics in group work, although not necessarily in the presence of conflict or tension. 

\subsection{\label{sec:framing}Epistemological framing}
 The physics education community has attended to implementation and impacts of group work, but we have thought less about how the \textit{students} think about group work, and how that interacts with their learning. Student perceptions of group work likely impact the roles or positions individual students have access to within a group~\cite{EddyEtal2015}. One way researchers examine students' perceptions or expectations of learning physics in a collaborative environment is through the lens of epistemological framing (e.g., ~\cite{Redish2004, ScherrHammer2009, IrvingMartinukSayre2013, ConlinEtal2007, DiniHammer2017, ModirThompsonSayre2017}). Generally, in the disciplines of anthropology and linguistics, framing refers to an individual or group's sense of what is going on in a given situation, including expectations about what could and should happen, what should be attended to, and how one should act~\cite{Tannen1993, Bateson1972, Goffman1986}. Epistemological framing specifically refers to a sense of what is taking place with respect to knowledge (e.g., Is this a situation for sense making, or for rote manipulation of formulas?); these frames can be considered for individuals or groups and can vary moment to moment~\cite{Redish2004, ScherrHammer2009}. There are other aspects of framing as well, such as social framing, which refers to expectations regarding individual and community roles and interactions in a social setting. In the physics education community we primarily attend to epistemological framing, as the nature of knowledge has particular importance in school settings, although it is noted that different aspects of framing can interact with one another, especially in collaborative settings~\cite{ScherrHammer2009}. 
 
Scherr and Hammer describe a resource-based account of epistemological framing in which a frame is a ``locally coherent pattern of activations''~\cite[p. 151]{ScherrHammer2009} of epistemological resources. As epistemological resources~\cite{HammerElby2002, HammerElby2003} may exist within an individual's mind or be distributed across a group of people, this resource-based account links two disparate approaches to framing---one with respect to individual reasoning and another with respect to social dynamics across groups~\cite{BrownHammer2008}. In a study of collaborative tutorial-style activities in an introductory physics class, Scherr and Hammer illustrate that verbal and nonverbal actions together provide evidence of students' epistemological framing and insight into the dynamics of this framing~\cite{ScherrHammer2009}. Irving, Martinuk, and Sayre also explore the dynamics of epistemological framing by looking at the transitions or shifts between frames~\cite{IrvingMartinukSayre2013}. They identified two axes along which discussions could be categorized---expansive versus narrow, and serious versus silly---and observed that the majority of frame shifts were initiated by the teaching assistant facilitating the collaborative learning situation. 

The notion of epistemological framing informs our analysis of the collective problem solving in a group of four students. Looking at our data through the lens of epistemological framing, we would infer a coherent epistemological frame, where the group has shared understanding around what kind of activity they are engaged in, evidenced by both their verbal and nonverbal actions~\footnote{Considering a global view of their overall behaviors, we characterize the students in the group as sharing epistemological frames, while noting that individual and group frames can and do vary moment to moment.}. Yet despite this alignment, we see one student in the group positioned as less knowledgeable. To understand this positioning, we identify an epistemological aspect of group work, and we associate students' behaviors with expectations as to how knowledge will be generated in the group (epistemic stances toward group work). We need this new tool to help us identify and describe different stances that individual group members might be taking regarding the role of group work in the construction and application of knowledge, despite a cohesive epistemological frame among the group regarding what kind of activity they are engaged in. Our analysis and discussion are distinct from epistemological framing because while framing can include group work as one of many factors, we focus on a finer grained analysis of just the epistemic stances toward group work and how these interact with other aspects of the social learning environment. As depicted in Fig. \ref{fig:venndiagram}, we consider these epistemic stances toward group work to be interactions between epistemological and social aspects of framing, where epistemological framing refers to “what is taking place with respect to knowledge”, while social framing refers to students’ “sense of what to expect of each other, of their instructor, and of themselves”~\cite[p. 149]{ScherrHammer2009}. Social framing does not inherently include ideas about knowledge generation, and epistemological framing does not inherently include ideas about social interactions. In attending to epistemic stances toward group work, we consider the overlap, or blending, of epistemological and social framing.   

\subsection{\label{sec:productive}What counts as productive group work?}
In much of the prior literature, group work is considered productive or effective if the group engages in sophisticated problem solving strategies or if they reach a correct answer~\cite{HellerHollabaugh1992, Barron2003}. We expand this consideration and attend to multiple aspects of productivity. If a group is making progress toward a correct or sophisticated conceptual understanding of the physics, we call that group work “conceptually productive”. Yet, as we will see in our analysis below, a conceptually productive group does not ensure equitable discourse. A second aspect of productivity is whether all students in the group have access to participation in a variety of roles. Using the language of Eddy \textit{et al.}~\cite{EddyEtal2015}, we would consider a group to be more productive if there are fewer barriers to participation for the individual group members. In our analysis, we foreground this latter view of productivity---a productive group is one that engages in equitable discourse---while noting that there are multiple factors that contribute to, and dimensions along which to view, the productivity of collaboration. 

\section{\label{sec:methods}Methodology}
The case study presented in this paper comes from a broader study of students’ mathematical sense making in quantum mechanics (QM)~\cite{DreyfusEtal2017, SohrGuptaElby2018, JohnsonSohrGupta2018, GiffordPERC2019}. In order to investigate the ways students engage in MSM, we conducted a series of  focus group studies with Modern Physics and QM students in which we gave them QM problems to work on in a group for one hour~\cite{GiffordPERC2019}. While watching a video of one particular group and looking for elements of mathematical sense making, we noticed that one student was being positioned as less knowledgeable despite guiding much of the group's sense making. This motivated us to focus on this group and to analyze the intersections between epistemology and social dynamics during the collective sense making process. We take an analytic case study approach in order to examine a puzzling phenomenon through observation, description, and interpretation of a situation in context, with the goal of producing a rich and in-depth understanding~\cite{Merriam1985}.

The group consists of four students---Penny, Morgan, Cam, and Sarah (pseudonyms)---from a modern physics class. We did not have relationships with these students prior to the one-time focus group session that took place in the last week of the semester. We recruited the students by sending an email to the class which framed the focus group study as an opportunity to work through QM problems with their peers before the final exam, but was not directly connected to their grade in the course. The professor of the course, however, offered extra credit to students who participated. We organized the groups based on scheduling constraints, and the students were monetarily compensated for their time. Penny (she) identifies as a white female, and is a sophomore majoring in chemistry and physics. Morgan (they) identifies racially as white and responded to the gender question on our demographic survey with a question mark. They are a junior computer science major. We use “they” pronouns for Morgan~\footnote{We regret that we did not ask the students which pronouns they prefer, and thus use ``they'' for Morgan given the ``?'' response to the gender field. We have now made it a norm to ask students for their pronouns when conducting focus group and interview studies.}, and note that their gender performance includes many attributes typically considered to be masculine~\cite{TraxlerEtal2016, Butler1988}; we believe this is important information for interpreting the social dynamics we observe among the group, particularly because Morgan comes to fill the role of an explainer and engages in masculine forms of discourse that are common and favored in the broader physics culture~\cite{WolfePowell2009, Tonso2006, HawkinsPower1999}, and in at least one moment other students in the group refer to Morgan using ``he’’ pronouns (line \ref{Che} in the transcript). Cam (he) identifies as a white male, and is a sophomore physics major. Sarah (she) identifies as a white female and is a sophomore majoring in astrophysics and creative writing. While we suspect that gender plays a role in the interactions among these four group members, we have not yet analyzed the specific ways in which the interactions are gendered. Here, we focus on the entanglement of epistemological and social aspects of the group problem solving, and we leave deeper analyses of gender and power dynamics for a future paper. Although this paper does not provide analysis or claims around gender, we include gender information here because we believe it is important context for understanding and interpreting the group’s interactions.

\begin{figure*}
\centering
\includegraphics[scale=.5]{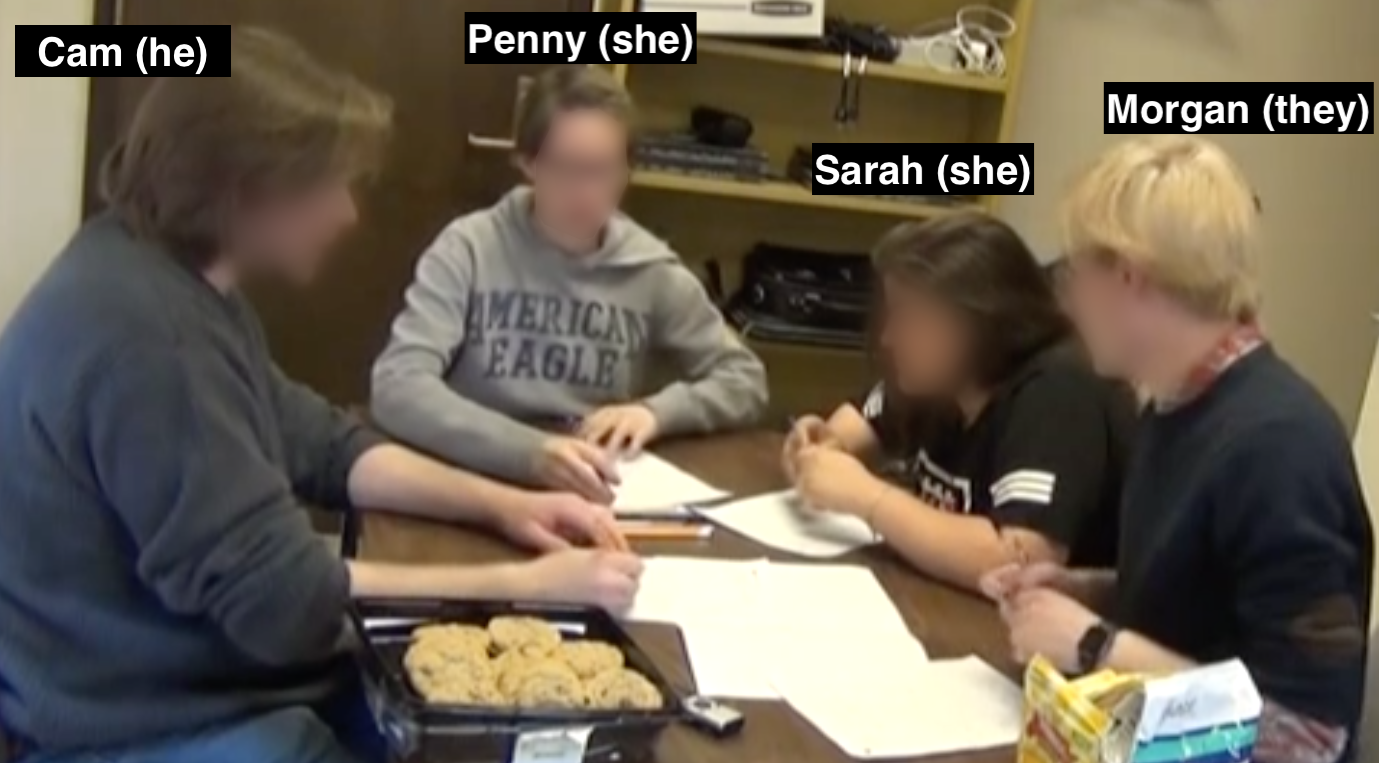}
\caption{Four modern physics students---Cam, Penny, Sarah, Morgan (starting on the left, going clockwise)---working collaboratively on QM problems.}\label{fig:group}
\end{figure*}

\begin{figure*}
\centering
\includegraphics[scale=.9, frame]{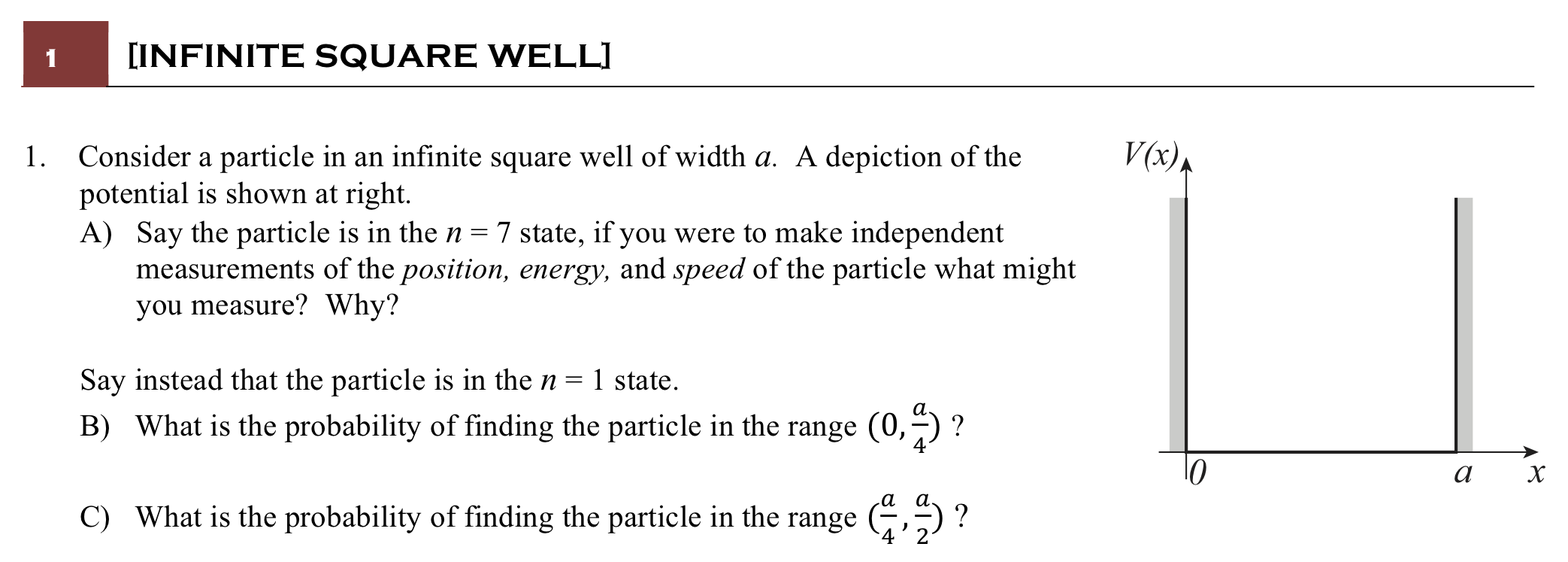}
\caption{Problem 1 from the focus group study asks about an infinite square well. In Episode 1, the students are discussing part A) and in Episode 2, they are determining how to check if their answers to parts B) and C) are correct.}\label{fig:Q1}
\end{figure*}

\begin{figure*}
\centering
\includegraphics[scale=.9, frame]{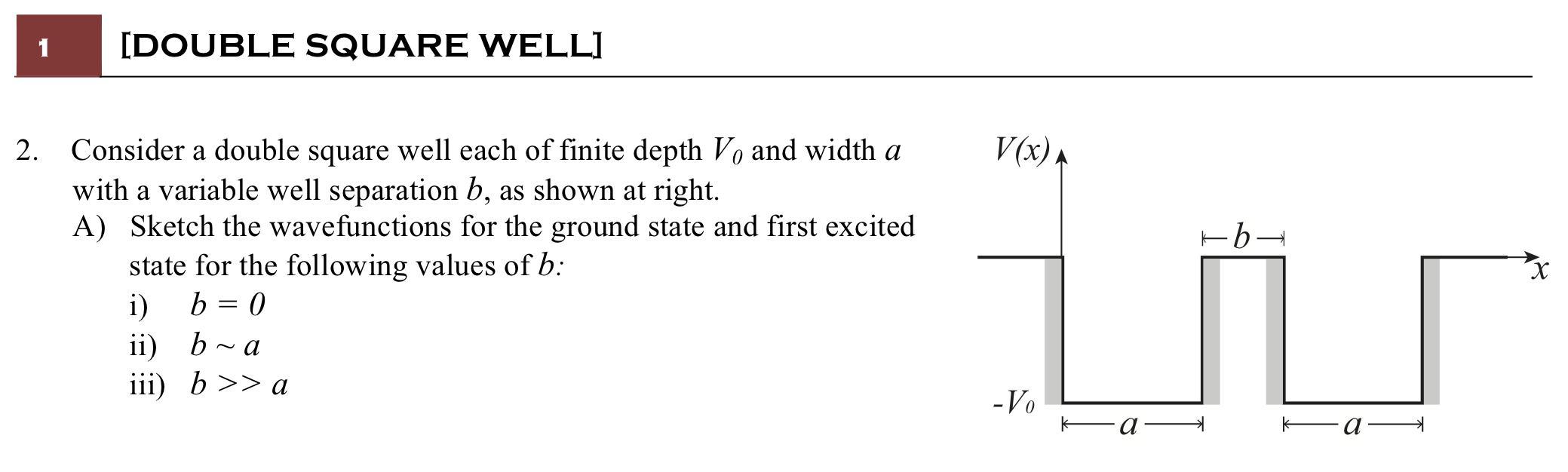}
\caption{Problem 2 from the focus group study asks about a double square well. In Episode 3, the students are discussing part A) ii) when $b\sim a$.}\label{fig:Q2}
\end{figure*}

The focus group took place in a small room with the students seated around a rectangular table---Morgan and Sarah sat on one side of the table, Cam on the other side, and Penny at the head of the table (see Fig. \ref{fig:group}). We framed the focus group session as a chance for the students to work on a few QM problems together. We asked them to talk to each other and say what they were thinking out loud as much as they could, and we noted that we were more interested in their sense making rather than them arriving at a correct answer. One researcher (the second author) was present in the corner of the room, and chimed in every once in a while with follow up questions for the group. We prepared four different problems (each with several sub-parts) for the students to work on; the question prompts were in line with QM content they had covered in their course. The three episodes we present here include the students working on two different problems. The first problem (see Fig. \ref{fig:Q1}) is about an infinite square well. Part A) asks the students to determine which values they might measure for position, energy, and speed for a particle in the n=7 state of the infinite square well. Parts B) and C) ask for the probabilities of finding the particle in the first and second quarters of the well. The second problem (see Fig. \ref{fig:Q2}) is about a double square well, and has the students consider the shape of the wave function when the well separation is: zero (i.e., one wider well), comparable to the width of the wells, and very large compared to the width of the wells.

We transcribed the hour-long video, and watched it several times from different lenses, attending to: the mathematical sense making students were engaging in (our original intention in conducting the focus groups), the interplay between social dynamics and sense making, the ways in which certain students get positioned as more or less knowledgeable, the inferred epistemic aspects of students' sense making, and the inferred epistemic stances toward group work. For individual episodes within the hour-long focus group, we conducted a discourse analysis~\cite{Sfard2001} in order to infer epistemic stances toward group work and explore the ways in which those interacted with the social dynamics and sense making to result in the positioning of students as more or less knowledgeable. We achieve an in-depth understanding of the group work scenario by attending to multiple aspects of the data---verbal interactions among the students (represented in the transcript), non-verbal interactions (captured in the transcript and in our narrative description), patterns of action and non-action (captured in the narrative description), and artifacts of the students’ written work (not included in this paper as a site of analysis, but we refer to them to determine who has control over the production of written work).   

We selected three focal episodes for this analysis---one from the very beginning of the hour-long session, one 17 minutes in, and the third 46 minutes in. We intentionally chose episodes from different times throughout the session when the students were solving and talking about different problems to get a sense for how the group’s interactions changed or stayed the same over the course of the hour. The episodes we selected contain rich conversation and interactions for which we could identify the intertwining of epistemological and social elements. However, these three episodes are not unique in this regard---many of the other episodes throughout the session contain similar interactions to those that we present here. Within each episode, our analysis looks at both the episode overall (a coarse grained view of what is happening, what they are talking about, the roles people are taking on, etc.) and a finer-grained look at the interactions among students (line by line interactions, turns of talk, gestures, etc.). There are two objects of focus of the present analysis that we first identify separately and then explore the connections between: epistemic stances toward group work, and social dynamics. In order to infer epistemic stances toward group work, we look for evidence of students’ expectations of how knowledge will be generated in a group. For example, if a student is leaning in to the group and looking at other group members we would infer that they were expecting to work together (versus an expectation of an individual activity). If a student is thinking out loud, and asking other group members to weigh in on their ideas, we infer that the student is enacting a stance that generating knowledge in a group involves collective co-construction~\cite{LeupenKephartHodges2020, Chi2009, PawlakIrvingCaballero2018} of ideas (versus an expectation of individual construction and subsequent sharing of ideas). Next, we attend to the social dynamics of the group by noting who is talking when and how the group members are attending to one another (or not); this includes identifying the words students are using, their body language, physical positioning, gestures, tone of voice, pitch and pace of speech, and who has control over the sense making the group engages in. Further, we ask how individual students are being positioned (by themselves and each other) within the group. The inferred epistemic stances toward group work and the social dynamics and positioning can be tightly intertwined with one another; looking for evidence of each independently helps us to explore the ways in which they are connected. 
 
The analysis focuses primarily on Penny, Cam, and Morgan; the fourth student, Sarah, is engaged and listening to the conversation (as evidenced by her posture leaning in to the group, smiling, nodding, looking at other group members, following along on her paper, and occasionally chiming in with “yeah”) but makes very few of her own verbal contributions. These types of contributions, which serve meta-conversational functions, are sometimes referred to as ``back channeling''~\cite{LeupenKephartHodges2020}. The social dynamics involving Sarah (including her silence) are worthy of a study in their own right, but they remain outside of the scope of the present paper. 

In the transcriptions, ellipses (...) represent pauses longer than those natural in speech; gestures or nonverbal actions are indicated in [square brackets]; square brackets sometimes also contain information added to the transcript by the researchers for clarity; em dashes (---) indicate interruptions in conversation or people talking over one another. In the data we present below, there are many instances of the students interrupting each other. When there is single pair of em dashes, the reader should read this as two consecutive turns in a conversation but with the first speaker getting cut off and yielding to the second speaker. For example,  
\begin{itemize}[noitemsep,topsep=0pt]
\item[]\textit{Penny: according to the probability density---
\item[]Morgan: ---But they'd conform to whatever density it is}
\end{itemize}
should be read as two turns in a conversation, where Penny's turn gets cut short and she stops talking as Morgan takes the floor. When there are multiple turns of talk, with each speaker's words book-ended with em dashes, this should be read as the individuals speaking over one other. For example, 
\begin{itemize}[noitemsep,topsep=0pt]
\item[] \textit{Cam: ---So basically like this---
\item[] Morgan: ---you draw that there's gonna be seven peaks---
\item[] Cam: ---And really---
\item[] Morgan: ---and nine nodes---
\item[] Cam: ---if you actually did the probability it'll actually like bounce up like---}
\end{itemize} 
should be read as Cam and Morgan each contributing their ideas to the conversation, but not listening and responding to one another. In this case, Morgan interrupts Cam's statement, but Cam continues to talk, and vice versa. This exchange should be read as quick, consecutive turns in a conversation, noting the interruptions and often simultaneity of multiple speakers' contributions. In a transcript of this nature, it may be helpful for the reader to read all of the statements from one speaker to get a sense for what they were trying to say while the other speaker was also talking (i.e., reading Cam's three statements contiguously, skipping over Morgan's statements). 
 
Over the course of two years, discussions among seven physics education researchers (the three authors along with collaborators from the University of Maryland Physics Education Research Group) led to many of the insights and preliminary versions of the arguments in this paper. Those arguments were then refined by the three authors, a team that includes two men and one woman~\footnote{In the analysis, we sought to remove our own views as much as possible, but note that we personally view group work in a way that is mostly aligned with how we perceive Penny's actions in the group (i.e., valuing externalization of confusion, and thinking through ideas in the moment with other people). We do not pass value judgements on which stance toward group work is ``right'' or ``better'', but note that our personal stances are generally aligned with Collective Consensus Building (although these stances are certainly context dependent) and that there are many other stances within the physics and physics education communities.}. As a woman in physics, JRH has privileged access to interpretation of the social cues and norms we see play out in the conversations among feminine and masculine performing physics students. This perspective thus shapes how we perceive and interpret the interactions among this particular group of students. Further, after reaching consensus among the research team, we validated and refined our analysis through additional discussions with physics education researchers external to the project. 

\section{\label{sec:data-analysis}Data and analysis}
In this section, we present three episodes with interleaving presentation of transcript and analysis.

\subsection{\label{sec:ep1}Episode 1}
Episode 1 takes place at the very beginning of the hour-long session. The students begin with question 1a (described above and shown in Fig. \ref{fig:group}). They are each looking down at their individual papers, reading the prompt. The conversation begins when Penny reads the question out loud and begins to share her thoughts:

\begin{itemize}[noitemsep,topsep=0pt]
\begin{linenumbers}
\item[] \textit{Penny: ``If you were to make independent measurements of the position, energy, and speed, what might you measure?'' Uhmmm.  \linelabel{Pidea} So since the, like at every, every time you measured you'd just like get a different position.  Right? \linelabel{Pright}
\item[]Morgan:---Yeah---\linelabel{Myeah}
\item[]Sarah:---Yeah---
\item[]Cam:---Mhmm---\linelabel{Cyeah}
\item[]Penny: Those would just, like according to the probability density---
\item[]Morgan: ---But they'd conform to whatever density it is, I'm not quite sure.\linelabel{Mnotsure}
\item[]Penny:	Like, whatever the, like the psi squared thing--- \linelabel{Pquestion} [writing on her paper]
\item[]Morgan: ---Yeah---
\item[]Cam: ---Yeah---
\item[] Sarah: ---Yeah---
\item[]Cam: ---That would represent the distribution.
\item[]Penny: And it would have to be zero at the edges so it would be like that. [drawing on her paper]
\item[]Morgan: And since it's the seventh excited state would there be---
\item[]Cam: ---Yeah, because we're infinite---
\item[]Morgan: ---seven peaks I think?\linelabel{Mthink}
\item[]Cam: Yeah
\item[]Morgan: It would be like--- [draws wavy sine peaks in the air with their pencil]
\item[]Sarah: ---Yeah}
\end{linenumbers}
\end{itemize}

From the beginning few seconds of the episode, we infer that all four students have the expectation that they are there to work together. This is evidenced by their posture, who responds to the questions, and the tentative nature of their questions. All four students are leaning in with their elbows on the table, looking down at their own papers but as the conversation begins they also look up at each other and each other's papers (see Fig. \ref{fig:group}). When Penny puts forth her idea in lines \ref{Pidea}-\ref{Pright} that you would measure a different position every time, and follows up with a question (``right?''), the three other group members respond simultaneously (``yeah,'' ``Mhmm,'' lines \ref{Myeah}-\ref{Cyeah}). Penny looks at Cam when she says ``right?,'' but he is looking down at his paper, as are Morgan and Sarah.  The fact that all three students respond, despite not being addressed directly or not looking at the person who asked the question, suggests an expectation that any idea put forth is for the group to make sense of and respond to. 

Additionally, to begin the hour-long problem solving session, the students put forth ideas with tentativeness, phrasing their ideas as questions (Penny lines \ref{Pright}, \ref{Pquestion}) and explicitly saying they're ``not quite sure'' (Morgan line \ref{Mnotsure}) or qualifying an idea with ``I think'' (Morgan line \ref{Mthink}). In the first few seconds of the episode, Penny and Morgan display this tentativeness, which sets the tone for a collaborative sense making environment. As the hour progresses, however, we see this dynamic change as Penny continues to be tentative while Morgan and Cam adopt more authoritative ways of engaging in the conversation. Cam continues the conversation by putting forth his own tentative idea:
\begin{itemize}[noitemsep,topsep=0pt]
\begin{linenumbers}
\item[] \textit{Cam: Well, your n would be seven times pi I believe, right?  Inside the k equals...
\item[] Morgan: Right.
\item[] Penny:	How does n change the number of like---\linelabel{Pn}
\item[] Cam: ---Uhmm because...\linelabel{Cint}
\item[] Morgan: So, I think---\linelabel{Mint}
\item[] Penny:	---waggle things---
\item[] Morgan: ---I think it's...
\item[] Cam: [writing on his paper]
\item[] Morgan: Right.  That's the, that's what's inside the sine function.
\item[] Penny:	Oh, like the sin(kx)
\item[] Cam: So it's 7 pi over a.
\item[] Morgan: That's what I thought, but I think what that...I think it means that if---
\item[] Cam: ---So basically like this [pointing to his paper]---
\item[] Morgan: ---you draw that there's gonna be seven peaks---
\item[] Cam: ---And really---
\item[] Morgan: ---and nine nodes---
\item[] Cam: ---if you actually did the probability it'll actually like bounce up like--- 
\item[] Morgan: ---Right---
\item[] Sarah: ---Yeah---
\item[] Cam: ---that many times---
\item[] Penny: ---Ohhh yeah---\linelabel{Pohyeah}
\item[] Morgan: ---'Cause in the second state it goes up once and then down---
\item[] Penny:	---It has that many nodes.  \linelabel{Pcontinues}Or does it have seven minus one or something?
\item[] Cam: I don't think we have a minus one here---\linelabel{Ctentative}
\item[] Morgan: ---I think there's a, I think it's a...Well because there's one peak, \linelabel{Mtentative}I think that what you've drawn [to Penny] is the, is the---
\item[] Penny:	---Yeah---
\item[] Morgan: ---the ground state---
\item[] Cam: ---n = 1---
\item[] Morgan: ---n = 1.  So that has one peak.  So I think it would make sense for it to equal the number of peaks.
\item[] Penny:	And like---
\item[] Morgan: ---Oh, \linelabel{Maha}it's n +1 uhh, zero points. That's what it is.}
\end{linenumbers}
\end{itemize} 

As the students try to figure out how many peaks the $n=7$ wave function has, Cam and Morgan begin to talk over Penny, and one another. Penny attempts to ask a question and join in the sense making, but gets interrupted at lines \ref{Pn} and \ref{Pohyeah}. At line \ref{Pcontinues}, she continues her thought that began with ``Ohhh yeah'' (line \ref{Pohyeah}), talking at the same time as Morgan and looking directly at Cam. She asks if the number of nodes is ``seven minus one or something,'' and Cam and Morgan both begin to answer her question. They do so tentatively, qualifying their explanations with ``I think'' (Morgan line \ref{Mtentative}) or ``I don't think'' (Cam line \ref{Ctentative}). This exchange represents a turning point in the overall dynamic of the group, where Penny begins to assume the role of ``question asker'' and Cam and Morgan begin to assume roles as ``explainers.'' These positioning and role-taking tendencies that we see developing in the first episode continue to evolve over the course of the hour. Here, Cam and Morgan explain to Penny as they try to figure out how the number of nodes depends on the state. This culminates when Morgan has an ``aha moment'' and states in an excited and authoritative manner that ``it's n+1...zero points. That's what it is'' (line \ref{Maha}). We note that in this instance we see Morgan's excitement around figuring out the answer (the ``aha moment'') coincide with their assumption of the role of explainer. Next, Penny steers the conversation in a different direction by bringing up a question about the graphs they are drawing on their paper (sketching the wave function on top of the potential energy graph given with the prompt):
\begin{itemize}[noitemsep,topsep=0pt]
\begin{linenumbers}
\item[] \textit{Penny:	Well, also it doesn't really make sense to draw this [wave function] on this [potential energy] diagram right? \linelabel{PQ1} Because, or wait, like what is the height of this? [gesture?]
\item[] Morgan: Well that...the height represents---\linelabel{Manswer}
\item[] Penny:	---Isn't that just like the number of---\linelabel{Ptalkover1}
\item[] Cam: ---The probability---
\item[] Penny:	---Yeah---
\item[] Morgan: ---It represents the probability---\linelabel{Mprob}
\item[] Sarah: ---Yeah---
\item[] Morgan: ---of finding it there.
\item[] Cam: Right, \linelabel{Cright} so that's where you're most likely to--- 
\item[] Penny:	\linelabel{PQ2}So it just doesn't belong...Like, what is V of x? Isn't that potential?
\item[] Morgan: Right. \linelabel{Mright}So that's why, that's why it's sorta---
\item[] Penny: ---So you wouldn't really draw this on this graph---\linelabel{Ptalkover2}
\item[] Morgan: ---not so great to draw there---
\item[] Penny: ---You'd just draw it like this [drawing a new graph]---\linelabel{Ptalkover3}
\item[] Morgan: ---because the axes aren't the same---
\item[] Penny:	---and this [y axis] would be like, probability---\linelabel{Ptalkover4}
\item[] Morgan: ---Exactly---
\item[] Penny:	---and then this [x axis] would still be position.
\item[] Morgan: \linelabel{Pend}And so that, drawing the probability curve makes sense, but yeah you're right on that it doesn't represent potential so it's, it's not the most meaningful curve there.\linelabel{Mend}}
\end{linenumbers}
\end{itemize}

Again, Penny is the one asking the questions (lines \ref{PQ1} and \ref{PQ2}) that Morgan and Cam take up by explaining to her. At line \ref{Manswer}, Morgan immediately takes up the question about drawing multiple graphs on the same axes; whereas in lines \ref{Ctentative}-\ref{Mtentative} Cam and Morgan's explanations were tentative, here Morgan answers in a more didactic manner. The tentative qualifiers (``I think'') are no longer present, and both Morgan and Cam contribute ideas with more conviction: ``Well...the height represents'' (line \ref{Manswer}), ``It represents the probability'' (line \ref{Mprob}), ``Right, so that's where...'' (line \ref{Cright}), ``Right. So that's why...'' (line \ref{Mright}). Amidst these explanations, Penny tries to chime in and join the sense making but gets talked over (lines \ref{Ptalkover1}, \ref{Ptalkover2}, \ref{Ptalkover3}, \ref{Ptalkover4}). From lines \ref{PQ2}-\ref{Pend}, Penny continues to verbalize her thought process despite Cam and Morgan focusing on their own work and talking over her.  

In this episode, Penny has control over the conversation in that she is the one asking questions or providing contributions that prompt the group to engage in sense making (considering what the wave function looks like, and then what it means to draw the wave function on the same axes as the potential energy). One might interpret Penny's questions as an indication that she is confused about the content, or at least more vocal about her confusion than the other students. While it is likely that some of her questions correspond to genuine confusion or tentativeness, we also observe that  Penny often frames ideas in the form of questions or bids \footnote{The tendency of women to frame ideas as questions is well documented in feminist discourse literature~\cite{Lakoff1973}. This, among other finer grained analyses of these interactions, will be the subject of a forthcoming paper.}, even when she may not be confused about the content. This is evident in that she often presents an idea followed by ``right?'' (lines \ref{Pright} and \ref{PQ1}). Determining whether Penny’s questions indicate confusion or tentative presentation of more certain ideas is not crucial for this analysis; the fact that her contributions are consistently taken up by the group as questions, or requests for explanation, serves to position her as less knowledgeable within the group.  Penny does not have agency in the conversation in the sense that she is continually interrupted and explained to in a didactic manner. However, she does have agency in the sense that her contributions (often taken up by the group as questions) drive the sense making and topics of conversation within the group. With the confluence of these interruptions and explanations we begin to see Penny as being positioned as less knowledgeable within the group despite the fact that her ideas or contributions are the ones guiding the flow of the conversation. 

We focus on Penny and Morgan in this episode and identify them as enacting two different epistemic stances toward group work. Penny's tentativeness and orientation toward inclusion are aligned with an epistemic stance that the way knowledge will be generated in the group is through collective consensus building or sense making. By adding ``right?'' to the end of her ideas, Penny invites others to contribute or weigh in on her ideas, consistent with a notion of collaboration that involves throwing out ideas for everyone to collectively grapple with and build on. Additionally, Penny proposes tentative ideas and thinks through them as she is talking. For example, in line \ref{Pohyeah} Penny begins a thought with ``Ohhhhh yeah'' and finishes it in line \ref{Pcontinues} (``It has that many nodes'') after Morgan inserts a statement. In line \ref{PQ1}, Penny brings up the idea of graphing multiple things on the same axes and interrupts herself to ask, ``Because, or wait, like what is the height of this?'' She continues by beginning to answer her own question, ``Isn't that just like the number of '' (line \ref{Ptalkover1}). This happens again in lines \ref{PQ2}-\ref{Pend}, where Penny's five contributions are all a continuous thought interspersed with Morgan's explanations. Penny's thinking through ideas as she says them is characterized by slower speech. Overall, this slower pace of speech and asking questions where her pitch goes up at the end convey a tone of tentativeness and inclusion; she presents her ideas with uncertainty and invites others to weigh in. We associate these actions and this tone with a view of group work as a collective consensus building process, i.e., the Collective Consensus Building stance. 

Morgan's actions on the other hand are aligned with a stance toward group work that it should involve individuals working to understand the ideas in order to explain to other group members. This is evidenced by their immediate responses to, or taking up of, Penny's questions with didactic explanations. In lines \ref{Cint}-\ref{Mint}, both Cam and Morgan interrupt Penny before she finishes asking her question. Morgan explains, tentatively still at this point in the episode, how n is related to the number of peaks or nodes in the wave function. In line \ref{Manswer}, Morgan immediately takes up Penny's question about the graph, and through line \ref{Mend} explains to her why it does not make sense to draw the wave function and potential on the same axes. In this exchange, Morgan's explanations come across as certain and authoritative, in particular in lines \ref{Manswer} (``Well...the height represents'') and \ref{Mright} (``Right. So that's why...''). Additionally, in line \ref{Mend} they validate Penny's original idea that superimposing the two graphs does not make sense. This validation recognizes Penny's contribution, while also placing Morgan in a position of authority. Morgan speaks at a quick pace, and their posture is directed toward Penny especially at the end of this episode when they point at Penny's paper while explaining. Morgan's fast speech, posture, immediate responses to questions, and didactic explanations convey a tone of authority and assuredness (primarily in the second half of Episode 1). We see these actions and this tone as an enactment of a stance that group work means one person with the desired knowledge will explain to others, i.e., the Explainer-Explainee stance. Cam engages in some of the same behaviors as Morgan (e.g., interrupting or immediately answering Penny), but it is less clear in this particular episode that his actions are aligned with the Explainer-Explainee stance. For clarity, we focus primarily on Penny and Morgan in identifying the two different stances in Episode 1. 

In Episode 1, we see the interactions between epistemology of group work and social dynamics begin to result in the positioning of Penny as less knowledgeable. Penny's slower speech means that Cam and Morgan have time to interrupt her, and her questions and tone of inclusion create space for Cam and Morgan to take up explainer roles. In particular, we see the interactions of these factors in the exchange between lines \ref{PQ2}-\ref{Pend}. Penny asks and answers her question all at once (``what is V(x)? Isn't that potential?''), and continues her thought through line \ref{Pend}. However, her slower speech and tentative framing of questions, in addition to Morgan's faster speech and tone of authority, result in Penny being interrupted, explained to, and positioned as less knowledgeable. 

In summary, Episode 1 sets the stage for Penny, Sarah, Morgan, and Cam to engage in group work around QM problems. The group is cohesive in that they seem to have a shared understanding that making sense of these problems will require both conceptual and mathematical reasoning, and that as the learners they will need to construct meaning using different representations (e.g., equations and graphs). We infer that all four students have the expectation that they will work together, yet as the episode unfolds we see individuals engaging in group work in different ways. Penny begins to take on the role of the question asker, while Cam and Morgan start assuming roles as explainers. Sarah is engaged and listening, and every once in a while affirms others' statements. We infer two different epistemic stances toward group work: Penny reflects Collective Consensus Building, and Morgan is aligned with Explainer-Explainee. In this first episode, we see these epistemic stances interact with the social dynamics of the group in a way that begins to position Penny as less knowledgeable despite the fact that her contributions are the ones driving the sense making that the group engages in. These roles and positioning continue to evolve and begin to solidify as the group continues to engage in collective problem solving. 

\subsection{\label{sec:ep2}Episode 2}
Following Episode 1, the students continue to discuss Question 1A, but when they get stuck wondering if the speed of the particle in the well is constant and if this violates the uncertainty principle, they decide to move on to part B which asks them to determine the probability of finding the particle in the first quarter of the well if it is in the ground state ($n=1$). The students immediately begin writing down an integral. They recall the ground state wave function, and integrate the square of the wave function from $0$ to $\frac{a}{4}$. They do the same thing for part C which asks for the probability of finding the particle in the second quarter of the well (this time, integrating from $\frac{a}{4}$ to $\frac{a}{2}$). Less than seventeen minutes in to the problem solving session, the group has finished the two integrals resulting in symbolic expressions for the answers to Question 1 parts B and C. Episode 2 begins when they are looking at their resulting answers and trying to figure out how to check if they are right. Penny begins the conversation by making a bid to check their answers against their intuition:

\begin{itemize}[noitemsep,topsep=0pt]
\begin{linenumbers}
\item[] \textit{Penny: Yeah, I don't really know, I dunno how to have a good intuition about---\linelabel{Pintuition}
\item[] Cam: ---Well one thing that does make sense---\linelabel{Ctry1}
\item[] Penny: ---\linelabel{Panswer}Well wait the probability should be less than one---
\item[] Morgan: ---Uhh, \linelabel{Mexplain}those,  those sum to one...that plus that plus that again plus that again [pointing to the paper in the middle of the table].
\item[] Cam: Well the originally... yeah---
\item[] Morgan: ---'Cause if it's symmetric around a over two
\item[] Cam: Yeah
\item[] Morgan: Then these two should add to get one half [pointing to paper]---
\item[] Morgan: ---And they do. \linelabel{Mtheydo}
\item[] Penny: Aah. \linelabel{Paah} Wait... a over---
\item[] Cam: Aaaah.  Yeah, they do! \linelabel{Cint1}
\item[] Penny: Wait, isn't--- 
\item[] Cam: ---That makes sense.\linelabel{Cint2}}
\end{linenumbers}
\end{itemize}

Although Penny's statement about intuition in line \ref{Pintuition} is not phrased as a question, functionally it serves as a question to the group, and is taken up as such. Penny herself answers it in line \ref{Panswer}, when she says ``Well wait the probability should be less than one.'' For the first few lines of this episode, Morgan was hunched over their paper doing math---they wrote down an expression that was the sum of their answers to parts B and C (twice), and found that they summed to one. They finish this math right as Penny suggests the probability should be less than one (line \ref{Panswer}) and look up and immediately begin explaining (line \ref{Mexplain}), pointing to the graphs the students had previously drawn on a big piece of paper in the center of the table. Here, we see Morgan again taking up the role of the explainer, speaking quickly and with certainty. They explain the symmetry of the wave function, with Cam chiming in affirmatively. Penny attempts to join in and make sense of what they are saying (out loud, in the moment), but gets cut off by Cam in lines \ref{Cint1} and \ref{Cint2} as he ``gets it'' and expresses excitement over understanding Morgan's explanation. Penny finally gets in a question and asks Morgan to explain again:
\begin{itemize}[noitemsep,topsep=0pt]
\begin{linenumbers}
\item[] \textit{Penny: Can you explain that again? \linelabel{Pexplain}
\item[] Morgan: So like, if we have, essentially this is our function \linelabel{Mfunction}right here.
\item[] Penny: Mhmm.
\item[] Morgan: And this divides it, a over two---\linelabel{Mdivides}
\item[] Penny: ---So we went---\linelabel{Ptry1}
\item[] Morgan:---is the halfway point---
\item[] Cam: ---Yeeahhh---
\item[] Morgan: ---so we basically, if you add these two together---
\item[] Cam: ---You should be at the halfway point---
\item[] Morgan: ---Then we should have this area---
\item[] Cam: ---Which is one half [gestures at Penny with a quick bouncing finger pointing motion]\linelabel{Cfinger}---
\item[] Morgan: ---Which should be one half---
\item[] Cam: ---Which happens---
\item[] Penny: ---Ohh---\linelabel{Ptry2}
\item[] Cam: ---'cause these cancel \linelabel{Ccancel}
\item[] Penny: Ohhhhh! [smiles and silently claps]---
\item[] Morgan: ---the 4 pi's over two cancel \linelabel{Mcancel}
\item[] Penny: Oh that's so!---\linelabel{Ptry3}
\item[] Cam: ---And another way---\linelabel{Ctry2}
\item[] Penny: ---And this is n = 1 so like, that is symmetric\linelabel{Psymm}---
\item[] Cam: ---And it makes more se---\linelabel{Ctry3}
\item[] Morgan: ---Yeah---
\item[] Penny: ---So we know that it's symme---\linelabel{Ptry4}
\item[] Morgan: And so, yeah}
\end{linenumbers}
\end{itemize}

Penny asks Morgan, ``Can you explain that again?'' (line \ref{Pexplain}), reinforcing her position as the question asker and Morgan's position as the explainer. Morgan immediately takes up the explanation and the following conversation is characterized by Cam and Morgan talking over one another while Penny tries to join in, but gets cut off (lines \ref{Ptry1}, \ref{Ptry2}, \ref{Ptry3}, \ref{Ptry4}). In this section (\ref{Pexplain}-\ref{Mcancel}), Morgan and Cam are explaining to Penny (in this case, she explicitly asked them to). Yet Penny still tries to join in and make sense of the argument along with them (e.g., ``So we went...'' in line \ref{Ptry1}). From the very beginning of the episode, Cam has been trying to contribute an idea about how they can know their answers make sense (``Well one thing that does make sense...'' line \ref{Ctry1}), but he keeps getting talked over (lines \ref{Ctry2}, \ref{Ctry3}). He finally contributes his thought as they finish up the conversation:
\begin{itemize}[noitemsep,topsep=0pt]
\begin{linenumbers}
\item[] \textit{Cam: You should be able to say too that it makes... B) makes sense because, B) being the first part of it, should be less---
\item[] Morgan: ---Should be less than the second part---\linelabel{Mfinish}
\item[] Cam: ---than the second part
\item[] Penny: Mhmm...
\item[] Cam: So if we're taking the difference---\linelabel{Cdiff}
\item[] Penny: ---Ohhh that makes sense!---\linelabel{Pgetsit}
\item[] Morgan: ---And that makes sense because...ya know, otherwise it'd be weird for it to sum to one.
\item[] Cam: Okay...
\item[] Penny:	Good job team---\linelabel{Pteam}
\item[] Cam: ---alright.  Well, at least it makes sense, or seems to make sense...Except I still don't understand what that first one is asking for. [laughs]}
\end{linenumbers}
\end{itemize}

Cam puts forth the idea that their answer makes sense because the integral for part B (first quarter of the well) should be less than that of part C (second quarter of the well) given the shape of the wave function, but Morgan interrupts to finish his sentence for him (line \ref{Mfinish}). When Morgan completes the explanation of the symmetry argument, Penny signals that she gets it by saying ``Ohhh that makes sense!'' (line \ref{Pgetsit}), and concludes the episode by saying ``good job team'' (line \ref{Pteam}). 

The majority of this episode consists of Morgan and Cam explaining to Penny. Penny has assumed the role of question prompter (line \ref{Pintuition}) or question asker (line \ref{Pexplain}), while Morgan and Cam have taken up the roles of ``knowers'' or explainers. These two different roles inform one another; for example, in line \ref{Pexplain} Penny asks ``Can you explain that again?'' which positions Morgan and then Cam (who take up the question) as explainers. We see Penny being positioned as less knowledgeable in this episode. Right off the bat, she prompts the question about connecting to intuition (line \ref{Pintuition}), and as she puts forth an answer, she gets interrupted by Cam (line \ref{Ctry1}) and then Morgan (line \ref{Mexplain}). Then when Morgan begins to explain the symmetry argument, Cam understands it first and his affirmative interjections (``Aaaah. Yeah, they do!'' line \ref{Cint1} and ``That makes sense'' line \ref{Cint2}) prevent Penny from being able to join the collective sense making process. She continues to be interrupted throughout the episode when she attempts to contribute to the sense making (lines \ref{Ptry1}, \ref{Ptry3}, \ref{Ptry4}). Building on the tendency we noticed in Episode 1 for Cam and Morgan to explain to Penny, Episode 2 is dominated by these didactic explanations (lines \ref{Mexplain}-\ref{Mtheydo} and \ref{Mfunction}-\ref{Mcancel}). Additionally, in line \ref{Cfinger}, as Cam joins Morgan in explaining to Penny that their answers from parts B and C should sum to one half, he gestures to Penny with a quick, bouncing finger-pointing motion. One might interpret this gesture as a manifestation of Cam's excitement for figuring out that their answer makes sense, or as a reflection of his position as knower and transmitter of knowledge to Penny. We think both of these interpretations (and likely others) can coincide. Regardless of the intent or emotion behind the finger-pointing gesture, it functions in the group as a symbol of the social position of Cam as a knower and explainer and Penny as a questioner or receiver of knowledge. 

We identify the students' actions in this episode to be again aligned with two different epistemic stances toward group work: Penny as aligned with a Collective Consensus Building stance and Morgan and Cam aligned with an Explainer-Explainee stance. Perhaps most indicative of these stances are the different ways that Penny and Morgan react to not being sure about something. As Episode 2 begins, Morgan is hunched over their paper working out some math, and once they add up their symbolic expressions and find that they sum to one, they look up and begin explaining to the other group members. This is one example of how when Morgan is not sure about something, they retreat into individual problem solving mode and when they figure it out they re-engage with group to explain their solution or idea to the other members of the group. We see this happen multiple times throughout the hour-long session; this tendency is aligned with an individualistic stance toward group work with the goal of understanding for yourself so that you can explain to others. When Penny is confused or unsure about something, she vocalizes her confusion, contributing ideas tentatively and thinking through them out loud in the moment. One example of this occurs at the beginning of Episode 2 when Penny makes a bid for considering intuition (line \ref{Pintuition}) and then continues to put forth an answer to her own question (line \ref{Panswer}). We see this tendency as being aligned with a stance that considers group work to be a process of collective sense making where individuals put forth ideas they are unsure about and others grapple with and build on them so that the group collectively makes sense of the idea in the moment.  

Penny's contributions are tentative, yet she continually tries to join in on collective sense making (lines \ref{Paah}, \ref{Ptry1}, \ref{Psymm}, \ref{Ptry4}), and praises the group's accomplishment (``Good job team'', line \ref{Pteam}). Through these actions, Penny conveys a tone of tentativeness, inclusion, and excitement to be a part of a team. Cam and Morgan on the other hand, through fast speech and immediately jumping in and talking over people (lines \ref{Mexplain}-\ref{Mtheydo}, \ref{Mdivides}-\ref{Mcancel}, \ref{Ctry3}-\ref{Cdiff}), didactic or authoritative explanations (lines \ref{Mexplain}, \ref{Mfunction}-\ref{Mcancel}), and pointing at Penny's paper or the common paper in the middle of the table in order to explain (lines \ref{Mexplain}, \ref{Ccancel}), convey a tone of authority and assuredness, positioning them as knowers or explainers. 

We identified these two different epistemic stances toward group work in the first episode and see them again in the second episode and throughout the hour-long session. We note, however, that each individual student does not always act in alignment with these stances or roles that we have identified. In line \ref{Pexplain}, Penny explicitly asks for an explanation from Morgan, which could be evidence of an Explainer-Explainee stance. However, a few lines later, Penny tries to join in on the explanation, which we take as a bid for collective consensus building. This moment-to-moment variation in inferred epistemic stances toward group work is in line with the idea that epistemologies can be context dependent or vary moment to moment~\cite{ElbyHammer2001, ChinnBucklandSamarapunga2011}. Likewise, there are instances where we see Cam and Morgan engaged in practices that suggest an orientation to collective sense making (e.g., tentativeness in the first half of Episode 1). We do not see these as contradictions to the overall interpretation, but instances of context dependence and fluidity of epistemic stances. Taking a global view of the hour-long session as a whole, we associate Penny's engagement with a stance of Collective Consensus Building, and Cam and Morgan's actions with an Explainer-Explainee stance. 

We have largely left Sarah out of our analysis, because she does not often contribute verbally to the group, which makes it more difficult for us to make inferences about her involvement in the group. There are two possible interpretations we might take from Sarah's silence. One is that her silence is indicative of an epistemic stance similar to that enacted by Cam and Morgan---that is, group work is a process in which one individual with knowledge explains to other individuals. If Sarah took this stance in this particular setting, and felt that she did \textit{not} have knowledge relevant to their problem solving, then she might be inclined to not make many verbal contributions, and would assume more of a role of a listener or consumer of the explanations provided by other group members. An alternative interpretation is that Sarah takes a stance more aligned with Penny---one of collective consensus building---but that she does not see herself as part of the collective group and thus she does not contribute her ideas or vocalize her confusions. Both of these interpretations (and likely many others) are plausible. For now, we continue to focus our analysis primarily on Penny, Morgan, and Cam for whom we have more verbal cues to analyze. 

In Episode 2, the interactions between the epistemic stances toward group work, social dynamics, and sense making result in Penny being positioned as a question asker with Cam and Morgan being positioned as knowers and explainers. Penny's slower speech and throwing out tentative ideas to think through on the fly, along with Morgan and Cam's faster and quick-to-explain speech, result in Penny being interrupted and explained to (e.g., lines \ref{Panswer}, \ref{Paah}). In this particular episode, these dynamics lead to Penny not contributing as much (in terms of turns of talk) as in other episodes. When Penny asks for an explanation (line \ref{Pexplain}), Morgan and Cam take it up with fast-paced and authoritative explanations that make it hard for Penny to contribute when she attempts to join in on the collective sense making (lines \ref{Paah}, \ref{Ptry1}, \ref{Psymm}, \ref{Ptry4}). Despite this social positioning, it is Penny's contributions that again guide the group in the sense making they engage in---she prompts them to consider intuition in line \ref{Pintuition} and symmetry in line \ref{Psymm}. We note that before this episode began, Morgan had already begun to check that their answers summed to one half, indicating that they were already considering symmetry and possibly intuition (although we suspect ``intuition'' might mean something different to Morgan than it does to Penny). However, if Penny had not directly asked about these aspects, we do not know if the group would have engaged in conversation around them in the way that they did in this episode.

\subsection{\label{sec:ep3}Episode 3}
Episode 3 takes place 46 minutes in to the hour-long session, when the students are working on Question 2A (see Fig. \ref{fig:Q2} above), sketching the ground state and first excited state wave functions for a double square well. They spent a small amount of time on part i), considering the situation in which $b=0$. In this episode, they are working on part ii) where $b\sim a$. The group has agreed that the wave function must be an exponential decay towards infinity on either end and a sine function within each well, but they are unsure what the solution inside the barrier (the middle region) should be. They have talked about an exponential decay coming from either well to meet in the middle, and this episode begins as they seek mathematical (formulaic) justification. They are trying to remember where the sine and exponential solutions come from in the first place, and Morgan begins to write down the Schrödinger equation on the large piece of paper in the center of the table:

\begin{itemize}[noitemsep,topsep=0pt]
\begin{linenumbers}
\item[] \textit{Morgan: So if we have a Schrödinger equation which looks essentially like negative---
\item[] Penny:	---hbar squared---
\item[] Morgan: ---a positive constant...
\item[] Cam: Yeah, okay---\linelabel{Caffirm1}
\item[] Morgan: ---times the second derivative of this [wave function]...uh...equals...
\item[] Cam: The E minus U of x---\linelabel{Chelp}
\item[] Morgan: ---E minus U of x
\item[] Cam: Times...Yuuuup---\linelabel{Caffirm2}
\item[] Morgan: ---times that [wave function] then when you rearrange this you get double prime of x equals...so if E minus U of x is negative, is less than zero, right this [$E-U(x)$ term] is negative this [$-K\psi''$] is negative, so we get...essentially the solutions to that differential equation. [circling $\psi''=+k\psi$ on the paper]}
\end{linenumbers}
\end{itemize}

From the beginning of the episode, Morgan assumes the role of ``math doer'' as they have control of the large piece of paper and thus the conversation. The other three group members are leaning in and all watching as Morgan writes down equations and explains out loud what they are doing. While Morgan explains, Cam chimes in with affirmations (lines \ref{Caffirm1}, \ref{Caffirm2}) that suggest he understands what Morgan is doing and also starts to help out with the explanation (line \ref{Chelp}). Penny continues the conversation by asking a question:

\begin{itemize}[noitemsep,topsep=0pt]
\begin{linenumbers}
\item[] \textit{Penny: \linelabel{Plocation}How does, uhm, the I guess like location along x determine whether E is like E minus U is less than zero or greater than zero.
\item[] Morgan: E is constant, but U of x is a function of x.\linelabel{Mexplaina}
\item[] Cam: Mhmm
\item[] Penny: Ohhh, yeah yeah yeah---
\item[] Cam: ---Yes. And so yeah, \linelabel{Cgraph}you could look at the potential---
\item[] Morgan: ---So as x changes this is gonna change. If E-U(x)---
\item[] Penny: \linelabel{Pjoin}Is gonna change from being zero to being V of---
\item[] Cam:---V naught---\linelabel{CV0}
\item[] Penny \& Cam: ---Yeah---
\item[] Morgan: \linelabel{Mmath}If it's greater than zero though we get this one with a negative sign there, that's a little hard to read. Uhm, and so the solutions of this are roughly...This is e to the x or e to the negative x. 
\item[] Penny: Yeah, that's usually---\linelabel{Pinta}
\item[] Morgan: ---And this is
\item[] Penny: What does that say? Psi...?
\item[] Morgan: Sorry. This? Yeah, I wrote that badly.  The double derivative of---
\item[] Penny: ---Equals negative psi? Okay.\linelabel{Pclarify}
\item[] Morgan: Yeah.  This leads to e to the i x which is the same as cosine x plus sine x---
\item[] Penny: ---Isn't it also just sine?---\linelabel{Psin}
\item[] Morgan: ---With some, ya know, there's \linelabel{Mconstants1} constants here.
\item[] Cam: Right. \linelabel{Cright2}
\item[] Penny: Just sine x?
\item[] Morgan: Hmm?
\item[] Penny: Doesn't sine x satisfy this?
\item[]Cam: Yeah. \linelabel{Che} But he's writing out the general form and then A or B would be zero---
\item[] Penny: ---Ohhhh. Yeah.---\linelabel{Punderstand}
\item[] Sarah: ---Yeah---\linelabel{Sunderstand}
\item[] Cam: ---depending on---
\item[] Morgan: ---Yeah---
\item[] Cam:---\linelabel{Cbegin}basically you'd model it to fit the well.  And you'd decide cosine or sine. Usually, in most cases, we'd use sine. But if it did need to start at one instead of zero then you would use cosine.\linelabel{Cend}}
\end{linenumbers}
\end{itemize}

In line \ref{Plocation}, Penny prompts the group to connect the equation Morgan has written down (particularly the E-U(x)) term to the graph, a move that we consider to be part of mathematical sense making~\cite{OddenRuss2019, DreyfusEtal2017, LenzEtal2017}. Morgan takes up this prompt and explains to Penny that the total energy is constant but the potential energy is a function of x, and Cam chimes in to refer to the graph (line \ref{Cgraph}). Penny then starts to join in the collective sense making (line \ref{Pjoin}), and Cam jumps in to finish her thought and help her out, saying ``V naught'' and pointing at the graph on her paper (line \ref{CV0}). Morgan returns to the differential equations and continues to work out the solutions, with Penny asking for clarification about what Morgan has written (lines \ref{Mmath}-\ref{Pclarify}). When Morgan says that the solution in the region where E-U(x) is positive is cos(x)+sin(x) (with some constants), Penny vocalizes a question---``Isn't it also just sine?'' (line \ref{Psin}). Once the group hears this question, Cam jumps in to explain to Penny what Morgan is doing (line \ref{Che}). He continues the explanation after both Penny and Sarah have indicated that they understand (``Ohhh yeah'', line \ref{Punderstand} and ``Yeah'', line \ref{Sunderstand}). Midway through Cam's explanation in lines \ref{Cbegin}-\ref{Cend}, Penny looks down as if she has stopped listening to the explanation. Morgan returns to the math on the paper and continues:

\begin{itemize}[noitemsep,topsep=0pt]
\begin{linenumbers}
\item[] \textit{Morgan: Right...Okay...So it does have to be the decaying one in the middle.  It \linelabel{Mcant}can't be a sine wave.  Well that's nice.\linelabel{Mnice}
\item[] Penny: Wait, why?
\item[] Morgan: Because \linelabel{Mbc}here we have the, we're in this condition [circling $\psi''=+k\psi$ on the paper], because E minus the potential, the potential is high up here [pointing to the graph on Penny's paper] and we presumably don't have the energy. \linelabel{Menergy}
\item[] Penny: Yeaahhh.  I mean...I don't know what E is.  What determines...So I guess \linelabel{Pguess}the thing that determines E is n.
\item[] Morgan: Right.
\item[] Penny: Through like [laughing] a relationship that we talked about already that we don't remember---
\item[] Cam: \linelabel{Cyou}---Right, yeah. What you could do is you could actually do your double derivative here with your you know your n value and your k stuff and you could actually solve for the energies and stuff if you needed.\linelabel{Cneed}
\item[] Morgan: Right...But like. That, this observation. So like when U of x is high [pointing to the graph on Penny's paper], above what we think the energy of our electron is...
\item[] Penny: Then...it---\linelabel{Pintb}
\item[] Morgan:---Then---
\item[] Penny: ---then it must be---\linelabel{Pintc}
\item[] Morgan:---we know that...uh psi of x has to take on a form that's a negative exponential. Or a positive exponential, but the positive exponential can't happen because of regularity conditions.
\item[] Penny: And then that's, that is intuitively right.\linelabel{Pintuitivecheck}
\item[] Morgan: Yeah.  \linelabel{Mconstants}There's a lot of constants here and stuff that I left out 'cause I don't actually know how to do ODE's---\linelabel{MODEs}
\item[] Penny: ---\linelabel{Presponsest}That if the energy is, if the potential energy is higher than the energy of the thing than the thing is not like as likely to be there.\linelabel{Presponseend}
\item[] Morgan: Right, well, yeah \linelabel{Mmodify}it drops off exponentially as you get further away from the uh low regions.}
\end{linenumbers}
\end{itemize}

In lines \ref{Mcant}-\ref{Mnice}, Morgan says, ``It can't be a sine wave. Well that's nice'' with certainty as if to say, ``Great, we have the answer!'' Penny begins to ask a question about the total energy but then interrupts herself and answers her own question, in a tentative way (``I guess'', line \ref{Pguess}). Earlier in the focus group session, the group recalled that the total energy (E) was determined by the value of n, but they could not remember the formula. Here, Cam jumps in to explain how they could get that formula (by solving the differential equation), although they do not actually pursue this. To conclude the episode, Penny prompts the group to check their answer against their intuition (line \ref{Pintuitivecheck}). Morgan takes up this ``intuition'' idea by considering mathematical formalisms, talking about the constants they left out because they ``don't actually know how to do ODE's'' (lines \ref{Mconstants}-\ref{MODEs}). Penny's response indicates that she meant something different by ``intuition''---``if the potential energy is higher than the energy of the thing than the thing is not as likely to be there'' (lines \ref{Presponsest}-\ref{Presponseend}). Morgan acknowledges Penny's idea, but modifies it (``well, yeah it drops off exponentially'' line \ref{Mmodify}), ending the episode in the didactic manner in which they started. 

In this episode, Penny again assumes the role of the question asker. As is characteristic of her contributions throughout the hour, she is particularly interested in connecting to intuition, and it seems as though she does not want to let her questions go until this intuitive aspect is satisfied. Morgan, in this episode, takes on the role of the ``math doer'' and explainer, controlling the shared piece of large paper and driving the mathematical argument that the group constructs in order to answer their question of what the wave function should look like in the barrier in between the two wells. Cam plays a slightly different role in this episode in that his contributions primarily validate, repeat, or add on to explanations. While he does not ``own'' or create the explanations, the act of validating what other people (usually Morgan) have said places him in a position of authority in the group. Sarah is quiet as usual, but is paying attention and leaning in to watch Morgan's orchestration of the algebra that leads them through the sense making process. 

We see Morgan has having most of the control over the conversation in this episode, yet Penny still opens up the group for sense making and guides the conversation with her questions. She prompts the group to connect their algebraic equation to the graph (line \ref{Plocation}) and to check their results against their intuition (line \ref{Pintuitivecheck}). As the question asker and prompter of these moments of sense making, we simultaneously see Penny being positioned as less knowledgeable within the group. Much like in the prior two episodes, Penny is interrupted (lines \ref{Pinta}, \ref{Pintb}, \ref{Pintc}) and explained to (lines \ref{Mexplaina}-\ref{Mconstants}, \ref{Che}-\ref{Cend}, \ref{Mbc}-\ref{Menergy}) throughout the episode.  

We again identify two different epistemic stances toward group work. Penny's tentative questions, ideas, and thinking out loud (lines \ref{Plocation}, \ref{Pguess}) and attempts to join in on the sense making process as Cam and Morgan explain to her (lines \ref{Pjoin}, \ref{Pinta}, \ref{Pintb}-\ref{Pintc}) are aligned with a Collective Consensus Building stance. Morgan's actions and contributions are consistent with an Explainer-Explainee stance, evidenced by their didactic explanations taking control of the large piece of paper and sometimes pointing at Penny's paper. Cam's actions are also aligned with this more individualistic stance toward group work, but they vary from Morgan's actions because Cam assumes more of a role of validator and explainer, as he continually validates, affirms, or repeats ideas (lines \ref{Caffirm1}, \ref{Caffirm2}, \ref{Cgraph}, \ref{CV0}, \ref{Cright2}) and explains what Morgan is doing or adds to their explanations (lines \ref{Che}-\ref{Cend}). Notably, in lines \ref{Cyou}-\ref{Cneed} when explaining how the group could figure out how exactly E depends on n, he uses ``you'' instead of ``we''. We interpret the ``you'' to not be directed at any one individual group member, but rather a replacement for ``one''. Had Cam used ``we'' instead, that would have reflected a different stance toward group work, perhaps suggesting he saw himself as part of a collective team, or that the knowledge could (and would) be generated in the group through a collective process involving everyone. The fact that he used ``you'' signals to us that he is not enacting an epistemic view of group work as a collective consensus building process. 

Episode 3 occurs almost at the end of the hour-long problem solving session. By this point, the tendencies and role-taking that we noted in Episode 1 have solidified as the students have learned or become comfortable in these particular roles. Throughout the hour, Penny has been interrupted, explained to, and positioned as less knowledgeable. Yet more than 45 minutes into the session, she continues to make space in the group for sense making, and is still attempting to contribute to a collective sense making process rather than just listening to Morgan and Cam's explanations (e.g., lines \ref{Pjoin}, \ref{Pintb}). The interactions between the different epistemic stances toward group work, the social dynamics, and the sense making processes are linked to the social positioning that we see develop in the group---Penny as less knowledgeable and a question asker, Morgan and Cam as knowers or explainers with positions of authority, and Sarah as a quiet listener who is not directly acknowledged by the other group members. 

\section{\label{sec:discussion}Discussion}
In our case study analysis, we characterize the group as being cohesive along some dimensions and not cohesive along others. The three most vocal group members appear to be on the same page about what counts as making sense of the QM problems at hand, and they engage in conceptually productive sense making \footnote{A common element of the group's problem solving, as initiated by Penny, includes valuing intuition as a tool for sense making, though we see possible variation within the group as to what counts as ``intuition.'' We do not explore this variation in the present analysis, though it would be an interesting site for future investigation}. We view this epistemological alignment in contrast to other groups of students in our data set (that are not included in the present analysis) in which we observe an epistemological conflict because one student wants to talk about the problems conceptually, while other students want to plug and chug through the mathematics and get answers to the problems. We view the sense making that the group engages in as conceptually productive because they demonstrate making progress toward content mastery \footnote{Making progress toward content mastery includes getting the right answer, but also understanding the concepts along the way well enough to be metacognitive, ask follow-up questions, check their answers, etc.}, engage in metacognition, find different ways to check their answers, and seek coherence among multiple representations. While the group members seem to be on the same page when it comes to epistemology of physics (what counts as learning physics or solving physics problems), we observe different orientations toward what it means to collaboratively generate knowledge (epistemology of group work).  Distinguishing between epistemology of physics and epistemic stances toward group work allows us to explore the connections between epistemology, group work, and social positioning. We argue that the group members enact different stances toward group work, and that this misalignment is one factor which informs, and is informed by, the social positioning of members within the group. 

We were motivated to conduct this analysis because we noticed a puzzling phenomenon in this particular group---the student who was steering most of the group's sense making (Penny) was simultaneously positioned as less knowledgeable within the group. Second to the prompts we gave them, the students' work is primarily guided by Penny's questions and bids for sense making. Throughout the hour, Penny largely assumes the role of the question asker or question prompter, doing most of the metacognitive work that moves the group's sense making along. Her interactions in the group are characterized by slower, tentative speech, and thinking through ideas out loud. Morgan and Cam assume roles as knowers and explainers, in many instances conveying a tone of authority and assuredness through their pace of speech, diction, posture, and gestures. There are two dimensions of epistemic agency at play here: Penny has agency in the sense that she is primarily controlling the topics of conversation, but Morgan, and at times Cam, have agency in the sense that they are the ones doing the talking and explaining. 

We have identified and inferred epistemic stances toward group work for individual members of the group, and see the dominance of one stance over the other manifest in the social positioning of students within the group. Time and again throughout the hour-long problem solving session, Penny tries to break in to the conversation with collective co-construction of meaning, but these moves are not taken up by the group (e.g., lines \ref{Ptry1}, \ref{Psymm}, \ref{Psin}). Instead, Cam and Morgan’s Explainer-Explainee stance emerges as the dominant approach toward collaboration in this setting (e.g., lines \ref{Mexplain}, \ref{Che}, \ref{Cbegin}). Penny is positioned as less knowledgeable within this group, as evidenced by interruptions, didactic explanations, and the group taking up most of her contributions as questions in need of explanation. Associated with this positioning, the dominant epistemic stance toward group work incorporates an Explainer-Explainee paradigm, a stance which is at odds with Penny’s stance of Collective Consensus Building. 

In this analysis, we have attended to both social dynamics among the group of students and social positioning of individual members relative to the group. When we refer to social dynamics, we are referring to the fine-grained interactional dynamics---interruptions, tone, body language, etc. Social positioning, on the other hand, refers to a broader story of the roles or positions that individual members play within the social activity and how those positions interact with or relate to one another. The existence of, and misalignment between, two different epistemic stances toward group work in this particular collaborative activity are intertwined with the fine-grained social dynamics. For example, Penny’s speech is slow, consistent with thinking out loud and contributing tentative ideas to a collective meaning-making process (Collective Consensus Building), while Morgan’s speech is fast and they have a tendency to jump in and explain their understanding to Penny (Explainer-Explainee). The confluence of these dynamics and enacted stances toward group work allow Penny to be interrupted and explained to often, which then perpetuates the interactional dynamics associated with the misalignment of stances. In this way, the epistemic stances and social dynamics mutually inform and feed off of one another to reinforce a particular social positioning. A result of this feedback loop is the positioning of Penny as less knowledgeable and Morgan and Cam as authority figures. 

There are some instances of unidirectionality  in the analysis, where we see the  misalignment of epistemic stances \textit{driving} the social positioning. For example, in Episode 1, we begin to see Penny positioned as less knowledgeable \textit{because} she offers tentative ideas and questions (aligned with Collective Consensus Building), which open the door for her to be explained to. However, in other instances, the directionality may go the other way around, with the social positioning helping to reinforce or crystallize the epistemic stances toward group work and associated social dynamics. That is, as the hour-long session progresses, Penny’s Collective Consensus Building stance and Cam and Morgan’s Explainer-Explainee stance may be reinforced \textit{because} Penny has been positioned as less knowledgeable (i.e., the positioning of Penny as a ``question asker'' and Morgan and Cam as ``explainers'' may provide more opportunities for Penny to be tentative and Morgan and Cam to be authoritative). We view the social dynamics and misalignment of epistemic stances  to be reflexively intertwined with the social positioning; they mutually inform one another as the interactions among students play out, resulting in a crystallizing of the dominant Explainer-Explainee stance while Penny is positioned as less knowledgeable as compared to Cam and Morgan. 

While this paper presents a case study analysis of one group, we have examples from other groups in our data set where we have seen this positioning happening (although it unfolds in different ways in different groups). We focus on one case study here to propose the construct of epistemic stances toward group work, and document how it can contribute to, and interact with, social positioning of students in collaborative problem solving environments. From the perspective of epistemological framing, we would conclude that this group has a shared framing of the type of activity they are engaged in---they are all leaning in to participate, they are engaged in sense making (as opposed to answer making)~\cite{OddenRuss2019, ChenIrvingSayre2013}, and they have some shared understanding that this sense making process will involve both conceptual understanding and doing math. However, focusing on only epistemological framing, we would miss out on the fine-grained social interactional dynamics that contribute to the positioning of individual students. The construct of epistemic stances toward group work helps us to identify and unpack these complicated interactions, which have implications for creating and fostering equitable group work environments.  

The sense making that this particular group engages in is beneficial for Penny, Morgan, and Cam in terms of making progress toward content mastery. However, there are other dimensions along which we might consider the group's interactions to be productive or unproductive. Considering who has access to certain roles or positions, we characterize this group as inequitable, and thus unproductive along the equity dimension~\cite{TheobaldEtal2017}. While Penny engages in scientific practices that we see as productive (metacognition, checking ideas against intuition, seeking coherence, etc.), she probably does not benefit from \textit{always} being the questioner. That is, Penny, and the group as a whole, would likely benefit from her sometimes assuming the role of an explainer. Likewise, we see Morgan and Cam engaged in sophisticated sense making processes and clearly demonstrating some level of content mastery, but we posit that these students could also benefit from engaging in practices like questioning, and checking ideas against intuition. Additionally, throughout the problem solving session, the group continues to come back to the Explainer-Explainee stance where Cam and Morgan are often explaining to Penny. Yet we imagine that it would be beneficial for the group as a whole (as well as individual members) to be able to play different kinds of games, sometimes incorporating an ``explainer'' and sometimes not. Some studies have shown that when one individual dominates the group work, individual members of the group demonstrate lower content mastery~\cite{TheobaldEtal2017, HellerHollabaugh1992}. Leupen, Kephart, and Hodges suggest that group work that is interactive and constructive~\cite{Chi2009} is beneficial for students' conceptual understanding because ``Constructive actions such as explaining or debating ideas and posing or answering questions involve students in generating new understandings and making meaning. When students are interactive as well as constructive, taking turns and building on one another’s thoughts, they draw on the power of socially mediated learning to prod each other along paths in their thinking that they would otherwise not take''~\cite[p. 2]{LeupenKephartHodges2020}. Theobald \textit{et al}, in a study of how group dynamics impact student learning of biology, also found that inequitable participation in group work settings helped to perpetuate social status of students in the group~\cite{TheobaldEtal2017}. Likewise, in our case study we see the social positioning of students within the group informing and being informed by the misalignment of students’ epistemic stances toward group work. Yet our analysis is more complicated than just a dominant individual or ``personality’’ hindering productive group work; while the dominance of Cam and Morgan’s Explainer-Explainee stance may be a barrier to equitable participation in the group~\cite{EddyEtal2015}, Penny is not just a ``timid’’ and ``silent record keeper’’~\cite{HellerHollabaugh1992}. Rather, Penny’s ideas drive the group’s sense making while Morgan does most of the math and Cam and Morgan do most of the explaining. In this way, each of the three students (Penny, Morgan, and Cam) have some form of epistemic agency. 

The interactions we observed among this group of students are in line with the documented gendered speech patterns and social discourse norms in science and engineering that favor masculine ways of engaging in conversation~\cite{WolfePowell2009, Tonso2006, HawkinsPower1999}. Motivated to understand the positioning of members within this group and the potential discounting of a female student who steers the group’s sense making, we zoom in on epistemological aspects of these social interactions. Although we do not directly address the gendered interactions in the present analysis, focusing on the intertwining of epistemology and social dynamics may provide one plausible mechanism for the gendered interactions and resulting social positioning. That is, investigating the interactions between epistemology and social dynamics gives us some insight into how this social positioning can happen in a collaborative problem solving setting. We attend to this epistemological aspect as a first step, and then in future work will dive more deeply into understanding the roles that gender, power dynamics, and ideology play in creating and driving interactions in this group problem solving setting.

\section{\label{sec:conclusions}Conclusions and implications}
We frequently ask students to work together in our physics classes, and while there are some instances where we attend to the group work by assigning roles or constructing groups based on gender or performance levels~\cite{HellerHollabaugh1992, GroverItoPark2017}, it is not a leading discourse in the physics education community to consider how to support students in engaging in group work productively and equitably. Through this exploratory analysis, we echo the calls of other researchers~\cite{PollockFinkelsteinKost2007, TurpenFinkelstein2010, Barron2003, Tanner2013, EddyEtal2015, PawlakIrvingCaballero2018, SohrGuptaElby2018} to not only implement group work in our physics classes, but to attend to the dynamics among students, investigate factors that contribute to collaborative environments, and construct group work settings that promote equity. Motivated by the observation that one student (Penny) continually steered the group's sense making and was simultaneously positioned as less knowledgeable, we investigated the interactions between social dynamics and epistemology. 

Through our analysis we see that it is not as simple as Morgan having more epistemic agency at the expense of Penny having less. Instead, we look at multiple dimensions of epistemic agency: who is controlling the topic, and who is doing most of the talking. We inferred students' epistemic stances toward group work---their stances about what it means to generate and apply knowledge in a group---and explored the ways in which the misalignment of these stances between individual students and the dominance of one stance over the other in the group as a whole contributed to, and was reinforced by, the social positioning of Penny as less knowledgeable.

The misalignment of the two different stances, along with the roles the students assume and the ways in which they engage in discourse, are intricately linked with the social positioning of Penny as less knowledgeable and Cam and Morgan as occupying positions of authority. We recognize that there are many factors at play in this social problem solving setting, including gender, power dynamics, and cultural norms and discourse patterns, and we identify the misalignment of epistemic stances toward group work (and the dominance of one over the other) as one factor that contributes to, and is also informed by, the social positioning of students within the group. 

We might consider possible reasons that the Explainer-Explainee stance dominates the group in this particular setting: Is it because there are more students in the group aligned with this particular stance (Cam and Morgan) than the other (Penny)? Or because the students aligned with this stance present as masculine? Or maybe it is because of the stance itself, as a more aggressive form of generating knowledge? Or perhaps because the more individualistic stance is more aligned with broader cultural norms in physics (or physics education)? Likely, most (or all) of these are at play. With the present analysis, we cannot answer these questions about why the Explainer-Explainee stance comes to dominate the group, but note that this is undoubtedly a complex phenomenon, and attending to epistemic stances toward group work is just one piece of the puzzle. The Collective Consensus Building and Explainer-Explainee stances embody different hierarchies; a forthcoming paper will include a complementary analysis of this same group that takes an even finer-grained look at how the local interactional dynamics reproduce, and derive from, broader patterns of discourse and cultural practices, and how power dynamics play a role in the interactions and social positioning~\cite{Erickson2004, Tonso2006}.

Understanding how multiple factors interact to privilege or exclude contributions from students in a group is a first step in learning how to cultivate, and support students in cultivating, equitable group work. The specific analysis we have presented is not meant to be generalizable to other groups of students or other contexts; every local situation will have different epistemological and social factors that contribute to how a group problem solving session might play out. However, the construct of epistemic stances toward group work can be applied in many situations within physics (or other discipline-based) education research. We have provided an example of how to attend to epistemic stances toward group work among a group of students, and demonstrated its utility as a tool for investigating interactions among students. Identifying students’ epistemic stances toward group work may be a helpful tool for instructors and researchers in working towards cultivating equitable group work. 

\begin{acknowledgments}
The authors would like to thank the anonymous students in this study and our University of Maryland collaborators---Andy Elby, Ayush Gupta, Erin Sohr, and Brandon Johnson---without whom this analysis and paper would not exist. This work is supported by NSF grant No’s 1548924, 1625824. Viewpoints expressed here are those of the authors and do not reflect the views of NSF.
\end{acknowledgments}

\bibliography{Epistemic_Stances_Toward_Group_Work}

\end{document}